# AN OVERVIEW OF VISIBLE LIGHT COMMUNICATION SYSTEMS

Taner Cevik[1] and Serdar Yilmaz[2]

[1]Department of Computer Engineering, Fatih University
Istanbul, Turkey

[2]Department of Electrical and Electronics Engineering, Fatih University
Istanbul, Turkey

## ABSTRACT

*Visible Light Communication (VLC) has gained great interest in the last decade due to the rapid developments in Light Emitting Diodes (LEDs) fabrication. Efficiency, durability and long life span of LEDs make them a promising residential lighting equipment as well as an alternative cheap and fast data transfer equipment. Appliance of visual light in data communication by means of LEDs has been densely searched in academia. In this paper, we explore the fundamentals and challenges of indoor VLC systems. Basics of optical transmission such as transmitter, receiver, and links are investigated. Moreover, characteristics of channel models in indoor VLC systems are identified and theoretical details about channel modelling are presented in detail.*

## 1.INTRODUCTION

Depending on the technological developments, the variety and quality of the communication devices and applications running on these devices have increased dramatically. These high quality applications require excessive data transfer capacity and speed.  Much of the internet transmission at the backbone is handled by the Optical Fiber Infrastructure that can achieve data speeds on the order of Tb/s. On the other hand, these high data rates at the backbone part cannot be perceived by the end users.  Nevertheless, it is not always beneficial and conceivable to deploy a cable infrastructure to every point of a site. Therefore, the importance of wireless communication increases day by day and is being widely used in the last-meter such as home, office and campus environments. Even though wireless communication is favorable in terms of cost, practicality and ease of operation, it brings about the bottleneck problem. RF waves that fall beneath the 10 GHz frequency portion of the electromagnetic spectrum have been widely used in wireless communication. However, since the existing bandwidth cannot satisfy the required capacity and speed demands, as well as multiple technologies contemporaneously share the same bandwidth (Wi-fi, bluetooth, cellular phone network, cordless phones), scientists and professionals have focused on new research areas in wireless communications. An alternative solution proposed for this first-meter bottleneck problem is shifting the working frequency interval to the unlicensed 60 GHz band. By this way, it is desired to widen the bandwith and achieve higher data rates [1].  Given the name WiGig, and standardized by Wireless Gigabit Alliance [2], it has become possible to reach about 6-7 Gb/s data rates with this technology [3].

However, shifting towards the right side of the frequency spectrum, reduces the wavelength of the electromagnetic waves. The propogation range of short wavelength signals is very limited. As the signal spreads over longer distances, the error rate increases due to the weakening of the energy [4]. Therefore, WiGig technology is intended to be used for data communication at high speeds in more enclosed areas.

Regarding to these quests, it is desired to utilize the mm-length electromagnetic waves ($\lambda <= 1mm$, $f > 100$ GHz) with the aim of enabling supplementary communication channels. Commuication with the mm wavelength on the right side of the spectrum is called Optical Wireless Communication (OWC). Data transfer on the infra-red band is already provided. Around 100 million electronic devices per year take place on the shelves adopted with infra-red technology. Moreover, the next generation wireless communication technology 4G and the follower are not built on a single technology. These Technologies are desired as an integrated top-one system that will compromise multiple technologies working in harmony. The OWC technology is expected to be an important figure of 4G and 5G systems especially in the section that the end users are connected to the internet [5]. The outstanding advantages of OWC when compared with Radio Frequency Communication (RF) can be listed as follows [6-10]:

- Unregulated 200 THz bandwith in the range of 155-700nm wavelengths.
- No licensing fee requirement
- Optical signals can not pass through walls like radio waves penetrate. Therefore, the signals emitted in a room provides significant benefits in terms of security by staying in that room. For long-distance communications Line-of-Sight (LoS) is essential, that is, the sender and the receiver must see each other directly. Any intervening situation or barrier can be easily recognized. Thus, OWC is significantly preferred in the military and state mechanisms that require high information privacy and security.
- Stay of signals in the room or office, eliminates the possibility of any interference in adjacent rooms or offices. By this way, each room will constitute a cell and the capacity prodcutivity will rise to the top levels.
- The equipment used is cheaper when compared with RF devices.
- Optical signals are not as detrimental as RF signals to the human health.
- OWC requires lower energy consumption than RF systems.

Data transfer by using the infra-red portion of the spectrum is already provided. Latest research activities have been focused on achieving data transfer simultaneously with enlightenment by means of using LED lighting equipment. These energy stingy and cost effective LED devices are desired to be used for data transfer without using RF signals, especially in short ranges. By using visible light, it is intended to achieve wireless communication in the environments and situations such as airplanes, hospitals etc, where it is not convenient to use RF waves.

The idea of illumination and data communication simultaneously by using the same physical carrier is firslty suggested by Nakagawa et al. in 2003 (Nakagawa Laboratory). Their studies [11-15] pioneereed many following research activities. Later on, the Nakagawa Laboratory went into coperation with the famous Japan technology firms and they established the Visual Light Communication Consortium (VLCC). Followingly, many research activities have been done that the most outstanding is the European OMEGA Project. Eventually, in 2011, IEEE completed the release and visual light wireless communication gained a global standard with the name 802.15.7-2011 [16]- IEEE Standard for Local and Metropolitan Area Networks--Part 15.7: Short-Range Wireless Optical Communication Using Visible LightUsing Visible Light [17]. Though a standard of visual light communication has been released in 2011, prevalent usage of this technology will take further time.

In this paper, we investigate the fundamentals and challenges of indoor VLC systems. In this paper, we explore the fundamentals and challenges of indoor VLC systems. Basics of optical transmission such as transmitter, receiver, and links are investigated. Moreover, characteristics of channel models in indoor VLC systems are identified and theoretical details about channel modelling are presented in detail.

The remainder of the paper is organized as follows. In section 2, we describe the fundamentals of VLC. Section 3 discusses the physical design of VLC links. Followingly, section 4 gives details about the channel modeling issues in VLC systems. Lastly, we give the concluding remarks and the future directions in section 5.

## 2. BASICS OF VLC

In recent years, one of the ideas put forward for wireless optical communication is the visible light communication method. The signals in the 380-780 nm wavelength interval of the electromagnetic spectrum are the light signals that can be detected by the human eye. It is possible to achieve illumination and data transfer simultaneously by means of LEDs that is the prominent lighting equipment lately. By this way, both interior lighting of a room and data transfer will be achieved without the need of an additional communication system. This technology is given the name of Visual Light Communication.

Basic configuration of a VLC communication system is given in Figure 1.

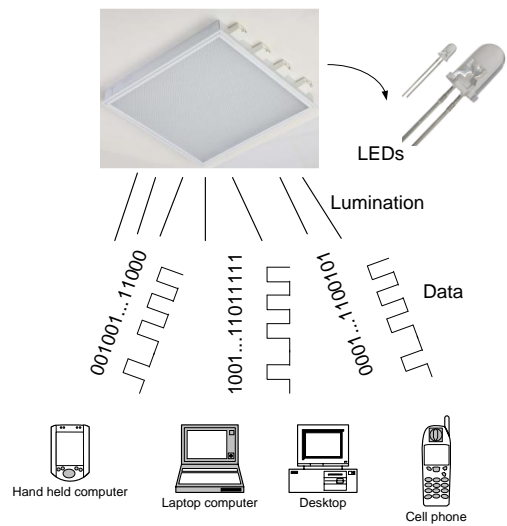

Figure 1. Basic VLC configuration

Fundamental entries in a VLC system are the transmitter (LEDs), receivers (photodetectors), modulation of data to optics and the optical communication channel as shown in Figure 2. We will discuss these main figures of a VLC system in the following section in detail.

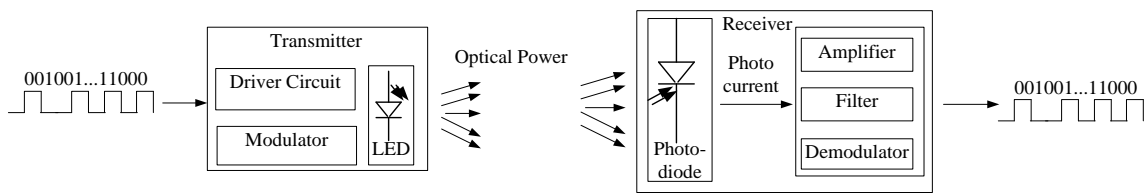

Figure 2. Block diagram of a VLC architecture

## 2.1. TRANSMITTER

There are different possible light sources used for illumination. However, Laser Diodes (LD) and LEDs are the two most popular ones among these especially preferred for optical data communication. Since the purpose of this study is about VLC, that is the concept of maintaining illumination and data transfer simultaneously, discussion of LDs is out of the scope. Thus, we will give details only about LEDs. The major difference between a LD and a LED is that, the former is a coherent light source and other is an incoherent source. That is, in the LED structure photons are emitted spontaneously with different phases. However, with LDs, a photon stimulates another photon that is radiated with a phase correlated with the previous which is called coherent radiation [10, 18].

### 2.1.1 BASİC PRİNCİPLES OF LEDS

The idea behind the emission of light with a p-n type LED is that, a bias voltage is applied to the p-n junction and by this way, holes in the p-junction move towards the opposite side. Also, the electrons residing in the n-junction are induced towards the p-junction. These minority carriers recombine in the depletion layer which is also called band-gap. In order for a high-energy level electron to recombine with a lower-energy proton, it should release the excessive energy. That is, the electron passes from the conduction band to the valence band. This excessive energy is released as a photon and approximately equal to the band-gap energy. The magnitude of the energy of the photon determines its wavelength which can be adjusted by the type of the semiconductor material. Relation between the band-gap energy and the wavelength of the emitted photon is given as follows [19]:

$$E_g = h * f \quad (1)$$

where $h$ denotes the Planck's constant with the value equal to $6.626 \times 10^{-34}$ Js and $f$ expresses the frequency of the radiated photon.

$$h * f = h * (c / \lambda) \quad (2)$$

where $c$ denotes the speed of light = $3 * 10^8$ m/s and $\lambda$ is the wavelength of the emitted photon.

#### 2.1.1.1 EFFİCİENCİES OF LEDS

An ideal LED should emit a photon per an injected electron. The ratio of the number of photons emitted to the number of electrons injected is defined as the internal quantum efficiency and represented in Eq. (1) [10, 20, 21].

$$\eta_{int} = n_{p\text{-}emt} / n_{e\text{-}inj} \quad (3)$$

where $\eta_{int}$, $n_{p\text{-}emt}$ and $n_{p\text{-}emt}$ denote the internal quantum efficiency, number of protons radiated from the active region and number of electrons injected respectively.

In an ideal LED, all of the photons in the active region, should leave te diode, which is called the unit extraction efficiency. However, not all of the photons emitted in the active region, leave the diode due to several reasons such as absorption, etc. So-called the extraction efficiency is another important parameter defining the efficiency of a LED. That is the ratio of the number of photons radiated into the free space per second to the number photons emitted in the active region per second and represented in Eq.(4) [20].

$$\eta_{extract} = n_{p\text{-}emt\text{-}air} / n_{p\text{-}emt} \quad (4)$$

where $\eta_{extract}$, $n_{p\text{-}emt\text{-}air}$ and $n_{p\text{-}emt}$ represent the extraction efficiency, number of photons emitted into the air and number of photons radiated in the active region respectively.

Consequently, we can derive the external quantum efficieny of a LED by combining the Equations (3-4) [22]:

$$\eta_{ext} = n_{p\text{-}emt\text{-}air} / n_{e\text{-}inj} \quad (5)$$

Ultimately, the power efficiency of a device is the ratio of the output power to the ratio of the input power, that is the electrical power applied to the LED ($P_{inj\_elec} = V * I$) to the emitted photon energy ($P_{photon}$).

$$\eta_{pow} = P_{photon} / (V * I) \quad (6)$$

### 2.1.1.2 RADİATİON PATTERN OF LEDS

A LED comprises of a semiconductor light source and a surrounding material with different refracting indexes respectively. The most popular LEDs are the ones with planar surfaces of which the emission pattern is modeled with Lambertian Law [23]. Although alternative surface shapes such as hemsipehere or parabolic are possible, fabrication of these LEDs are much more complicated. Light emerging from the semiconductor light source, faces with the surrounding material and refracts into the air with an angle different from the incoming one which can be explained by the Snell Law (Figure 3):

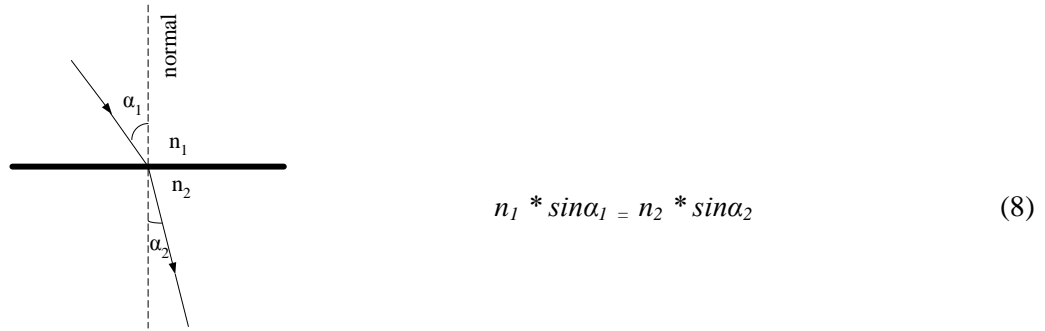

$$n_1 * sin\alpha_1 = n_2 * sin\alpha_2 \quad (8)$$

Figure 3. Refraction of a signal during the transition between two different materials with different refraction indexes

This situation decreases the external power efficiency of the device. If the signal incomes with an angle larger than the threshold, most of the signal will be reflected which reduces the external power efficiency to a fw percent [24].

Deriving from the Snell Law, a point directly on the direction of the light source with an angle of 0, gets the maximum intensity of the flux ($I_0$). However, at the point with an angle of $\theta$, the luminance intensity ($I_\theta$) is calculated as:

$$I_\theta = (P_{source} / (4\pi r^2)) * cos\theta \quad (9)$$

Since the air and semiconductor have different refractive indexes, the light intensity in air is derived as follows [20]:

$$I_{air} = (P_{source} / (4\pi r^2)) * cos\theta ((n_1^2) / (n_2^2)) \quad (10)$$

The above equations are valid for the assumption that the surrounding materila is hemisphere shaped and $I_{air}$ is the light intensity of the single point on the surrounding hemisphere. Hence, the total power power emitted into the air can be calculated as integrating the total intensities on the surface of the hemispehere as follows [20]:

$$P_{air} = \int_{\theta=0°}^{90°} I_{air} * 2\pi r * sin\theta * r * d\theta \qquad (11)$$

## 2.2. RECEIVER

Photodetectors are the receiver entity of an OWC system that absorbs the photons impinging on its frontend surface and overagainst generates an electrical signal. The conversion of photonic energy to the electrical energy can be achieved in alternative way. For example, in vacuum photodiodes or photomultipliers, the absorption of photons created photoelectric effects and free electrons emerge as a result that are used as carriers. Another way is that, by the falling of the photons into the junction area of a semiconductor photodiode such as p or pin diodes, an electron and hole pair is released. Followingly, these released carriers move to the corresponding regions such as conductance and valance bands in order to release their excessive energies [10, 25].

There are many types of photodetectors exist such as photomultipliers, photoconductors, phototransistors, and photodiodes that owning specific qualities. However, photodiodes are the most preferred devices as a photodetector due to their small size, high sensivity and fast response. P-I-N (PIN) and Avalanche Photo-Diode (APD) are the favored types photodiodes as a photodetector [26].

There are some important requirements that an ideal photodetector should cover:

- sensitive to the wavelength interval associated
- long operational life
- minimally affected from the temperature fluctuations
- efficient accomplishment of noise such ambient, dark, etc.
- noiseless physical structure
- small in size
- reliable
- cost effective

### 2.2.1 QUANTUM EFFİCİENCY

Different definitons are given for Quantum Efficiency. One of the definitions is that it is the probability of a single inicident photon to generate an electron-hole pair that contribute to the detection of electric current. Another definition for quantum efficiency is that it is the ratio of electron-hole pair generated that contribute to the detector current, to the incident photon flux and calculated as follows [27]:

$$\eta_{quantum\text{-}eff} = (1- R) * \xi * (1-e^{-ad}) \qquad (12)$$

where $R$ denotes the reflectance of the surface and that can be reduced by non-reflective coatings. $\xi$ represents the fraction of electron-hole pairs that contribute to the photocurrent. $a$ and $d$ exzpress the absorption coefficient (per $cm^2$) and the photodetector depth respectively.

### 2.2.2 RESPONSİVİTY

The responsivity of a photodetector is the ratio of the output current to the input optical power and the relation of it to the quantum efficiency is as follows [28]:

$$R = \frac{e*\eta}{h*v} \quad (13)$$

where *e* denotes the electron charge, h is the Planck's constant and v represents the light frequency.

## 3. PHYSICAL DESIGN OF VLC LINKS

One of the main challenges to be carefully considered during the design and modelling stages of an VLC system is the localization status of the transmitter and receiver pair which mainly defines how the signal is transmitted. Design of a VLC link can be classified in two ways as depicted in Figure 4. The first method of the categorization can be made whether the transmitter and receiver is directed or not to a specific point or coordinate. Under this category, three different options are possible regarding to direcitonality of the transmitter and receiver. The first option is that, both the transmitter and receiver are directed to a specific point. This type of configuration enhances power efficiency as well as immunity to the environmental distorting effects such as ambient and artificial light sources. The second category under the directionality classification is the nondirectional configurations in which the transmitter and receiver are not particularly focused to a specific direction or point. In order to achieve signal transmission, wide beam transmitters and wide FOV receivers are required. The main drawbacks of this configuration is the need for high power levels to combat with the high optical loss and the multipath-induced distortions. Although, multipath fading is overcomed by means of the immense ratio of the detector size and signal wavelength. The other link configuration of this type of classification is the hybrid design method in which the tranmitter and receiver can have different levels of directionalities, such as a narrow beam transmitter directed to a specific point and a wide FOV receiver which are not aligned to a particular direction [10, 29].

The second type of design choice is the existence of a LOS path between the transmitter-receiver pair. There are two options in this category of configuration. First is the Line of Sight (LOS) configuration in which no interruption or obstacles are present between the transmitter and receiver. Thus, no reflection consideration is considered that simples the path loss calculation. Besides, high power efficiency is achievable. In contrast, in the Non-LOS architectures, signals emerge from the source do not directly arrive at the receiver. They are refleected from different surfaces or objects and arrive in different time intervals to the receiver. This causes multipath distortions and maket he estimation of path loss much more difficult. The Non-LOS architecture with nondirected transmitter-receiver pair is called diffuse systems which is the most robust system and easy to implement for especially the mobile communication scenarios [10, 29, 30].

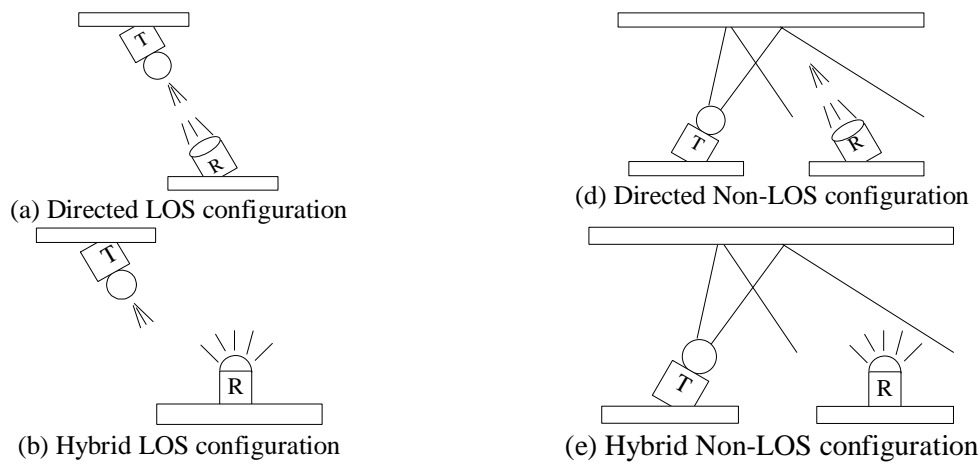

(a) Directed LOS configuration

(b) Hybrid LOS configuration

(d) Directed Non-LOS configuration

(e) Hybrid Non-LOS configuration

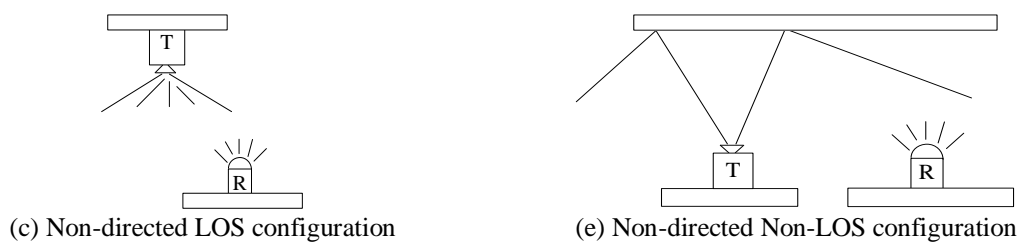

(c) Non-directed LOS configuration    (e) Non-directed Non-LOS configuration

Figure 4. Channel configuration models

## 4. CHANNEL MODELING

The most practical, cost effective and easy modulatin technique in VLC systems is the Intensity Modulation (IM) / Direct Detection (DD) [29, 31] . Unlike the coherent transmission techniques IM/DD does not concern with the nature of the signal such as phase or frequency. In contrast, in coherent transmission technology, information is coded on the optical beam by means of phase, frequency or amplitude modulation techniques. The receiver side owns the staff called down-converter which is comprised of an oscillator and a mixer. Mixer combines both the arriving optical beam and the one generated by the local oscillator. Then the combined optical signal falls onto the photodetector and depending on the similarity of the oscillator and incoming signal frequencies, either demodulation operation is applied or not. If the frequencies of the incoming signal and the oscillator are different, then a modulation operation is performed which is called Heterodyne Detection. Otherwise, the combined signal is directly downconverted to the baseband signal that is called Homodyne Detection [32].

For optical wireless systems, in which the system cost and complexities are desired to be low, IM/DD modulation technique takes over as a prominent method. In this technology, the desired waveform is modulated onto the instantaneous power of the carrier. At the receiver side, the detector uses the down-conversion technique DD, during which a photocurrent is produced directly proportional with the incoming photonic power [33].

Since the photodetector area is larger in the order of magnitudes when compared with the signal wavelength, VLC links do not suffer the impacts of multipath fading. However, since it is possible for signals reflected from ceilings, walls or other intervening reflective objects also arrive at the receiver by travelling in a dispersive way that results with the event multipath distorsion called InterSymbol Interference (ISI). For indoor VLC systems, this dispersion with ISI effect can be modeled as baseband linear impulse response ($h(t)$). This multipath distortion effect is mostly concerned especially for non-directional and Non-LOS channel models. Indoor VLC channels are generally assumed as quasi-static. In doing this, the positions of the tranmitter, receiver and the intervening objects are assumed to be static or moving with very low speeds and bit rate is very high. Thus, channel variations occur in the order of many bits periods [10, 29, 34]. The channel model for indoor VLC systems is based on the Eq. (14):

$$y(t) = Rx(t) \otimes h(t) + n(t) \qquad (14)$$

where *R* denotes the responsivity of the photodetector and formulated as in Eq. (13). $n(t)$ represents the gaussian modeled noise comprising ambient and preamplifier receiver noises. $\otimes$ expresses the convolution operation and $x(t)$ denotes the instantaneous input power which can not be negative ($x(t)>0$). Calculation of the average power received by the photodetector is as follows [29]:

$$P_r = H(0)P_t \qquad (15)$$

where $P_t$ denotes the average trasmitted power and calculated as in Eq. (16):

$$\lim_{T \to \infty} \left( \frac{1}{2T} \int_{-T}^{T} x(t) \, dt \right) \quad (16)$$

and $H(0)$ represents the channel dc gain that is defined as follows:

$$H(0) = \int_{-\infty}^{+\infty} h(t) dt \quad (17)$$

The channel modelling of indoor VLC systems differs according to the LOS presence between the transmitter and receiver. Following sections give details about the different models applied for the LOS and Non-LOS indoor VLC channels.

## 4.1 LOS CHANNEL MODEL

In an indoor VLC system, channel dc gain for LOS propogation model as depicted in Figure 5, with a source assumued as radiating in Lambertian model, concentrator with a gain of $g(\psi)$ at the receiver and an optical bandpass filter with the function $T_s(\psi)$ is given as follows:

$$H_{LOS(0)} = \begin{cases} A_r(m_1+1)\,(2\pi d^2)\,(\cos^{m_1}\phi)(T_s(\psi))(g(\psi))(\cos(\psi)), & 0 \le \psi \le \psi_c \\ 0, & \text{otherwise} \end{cases} \quad (18)$$

where $A_r$ is the effective area of the detector collecting the incoming optical signals with the arrival angle ($\psi$) smaller than the FOV angle that is expressed with $\psi_c$. $m_1$ is related with the directivity of the source beam and denotes the Lambert's mode number. In order to increase the efficiency of the system, transmitted optical power level should be increased. However, that is not possible because of the eye-safety and power efficiency reasons. An alternative solution can come into mind to employ larger dtector areas in the receiver. Although it can be thought as a promising solution, due to reasons of incereased cost, complexity, capacitance and reduced bandwith, this choice is unfavourable. The way of increasing the efficiency is using a concentrator at the frontend as mentioned before. $g(\psi)$ denotes the gain of the concentrator used at the front-end of the detector and calculated as follows [10, 29, 35]:

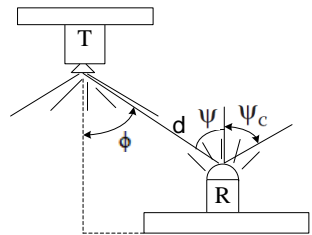

Figure 5. LOS channel configuration

## 4.2 NON-LOS CHANNEL MODEL

Channel modelling varies depending on the presence and degree of directivity of the transmitter and receiver. Althougy there are three types of Non-LOS configurations, there are two model generalizations covering the all. A single model can be used for the directed and hybrid Non-LOS configurations (Figure 6) as follows:

$$H_{Non\text{-}LOS(0)} = \begin{matrix}(\rho A h\, T_s(\psi)g(\psi)cos(\psi))/(\pi(h^2+d_{sr}^2)^{3/2}, & 0 \leq \psi \leq \psi_c \\ 0, & \theta > \psi_c\end{matrix} \quad (19)$$

where the parameter $\rho$ denotes the Lambertian reflectivity index of the ceiling. As clarified in Eq. (19), the best way of increasing the dc channel gain can be achieved by increasing the concentrator gain of the receiver by means of increasing its refractive gain and decreasing FOV angle [29].

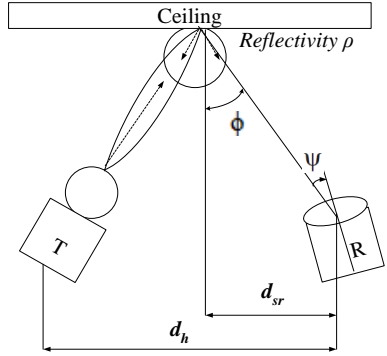

Figure 6. Directed or Hybrid Non-LOS Channel Configuration.

In non-directed non-LOS link configurations, high order reflections of the light emerging from the transmitter must be considered. Especially for the indoor environments such as with the higher than 5m in height, it is strongly advised to consider the reflections up to fifth order [36]. However, especially for small offices, the approach of considering the one-bounce reflection model estimates the dc channel gain with an error of a few decibels. First oder calculation of dc channel gain for the link configuration as depicted in Figure 7 is given in Eq. (20).

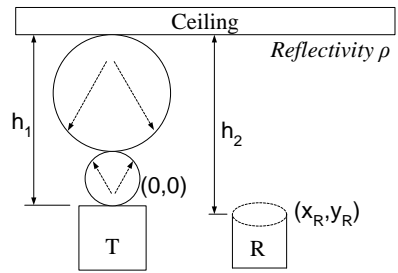

Figure 6. Directed or Hybrid Non-LOS Channel Configuration.

$$H_{Non\text{-}LOS(0)} = ((\rho T_s\, g A h_1^2 h_2^2)/\pi^2) \iint (h_1^2+x^2+y^2)^2 (h_2^2+(x-x_R)^2+(y-y_R)^2)^2 \quad (20)$$

where $T_s \approx T_s(\psi)$ since the optical filter is assumed as if it is omnidirectional and the concentrator is also omnidirectional with a gain $g \approx g(\psi)$, FOV $\psi_c \approx \pi/2$. Obviously, dc channel gain $H(0)$ is inversely proportional with $dh^{-4}$ [29].

## 5. CONCLUSION

Much of the internet transmission at the backbone is handled by the Optical Fiber Infrastructure that can achieve data speeds on the order of Tb/s. On the other hand, these high data rates at the backbone part cannot be perceived by the end users. Since the existing bandwidth cannot satisfy the required capacity and speed demands, as well as multiple technologies contemporaneously share the same bandwidth (Wi-fi, bluetooth, cellular phone network, cordless phones), scientists and professionals have focused on new research areas in wireless communications. An alternative solution proposed for this first-meter bottleneck problem is shifting the working frequency interval to the unlicensed 60 GHz band. By this way, it is desired to widen the bandwith and achieve higher data rates. Regarding to these quests, it is desired to utilize the mm-length electromagnetic waves ($\lambda <= 1mm$, $f > 100$ GHz) with the aim of enabling supplementary communication channels. In recent years, one of the ideas put forward for wireless optical communication is the visible light communication method. It is possible to achieve illumination and data transfer simultaneously by means of LEDs that is the prominent lighting equipment lately. By this way, both interior lighting of a room and data transfer will be achieved without the need of an additional communication system. This paper explores the fundamental issues and concepts of VLC systems by supporting theoratical details.

**AUTHORS**

**Taner Cevik** received the B.S., M.S. and Ph.D. degrees in computer engineering from Istanbul Technical University in 2001, Fatih University in 2008, and Istanbul University in 2012 respectively. In 2006, he joined the Department of Computer Engineering, Fatih University, as a research assistant, and in 2010 became an instructor at the same university. Since 2013, he has served as an assistant professor at Fatih University. 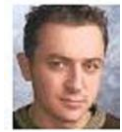

**Serdar Yilmaz** received the B.S., M.S. and Ph.D. degrees in electronics engineering from Istanbul University in 2000, Fatih University in 2003, and Yildiz Technical University in 2012 respectively. In 2000, he joined the Department of Electronics Engineering, Fatih University, as a research assistant, and in 2012 became an instructor at the same university. Since 2014, he has served as an assistant professor at Fatih University. 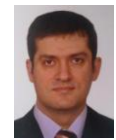